\documentclass[linenumbers, twocolumn]{aastex631}

\usepackage{subfigure}
\usepackage{url}
\usepackage{hyperref}
\usepackage{epsfig}
\usepackage{graphicx}
\usepackage{sidecap}
\usepackage{hanging}
\usepackage{color}
\usepackage{multirow}
\usepackage{enumerate}
\usepackage{natbib}
\usepackage{amssymb}
\usepackage{amsmath}
\usepackage[bottom]{footmisc}
\usepackage{apjfonts}
\usepackage{xspace}

\providecommand{\e}[1]{\ensuremath{\, \times \, 10^{#1}}}

\newcommand{\toi}{{TIC~393818343.01}}
\newcommand{\planet}{{TIC~393818343~b}}
\newcommand{\host}{{TIC~393818343 }}

\providecommand{\bjdtdb}{\ensuremath{\rm {BJD_{TDB}}}}

\providecommand{\msun}{\ensuremath{\,M_\Sun}}
\providecommand{\rsun}{\ensuremath{\,R_\Sun}}
\providecommand{\lsun}{\ensuremath{\,L_\Sun}}
\providecommand{\mj}{\ensuremath{\,M_{\rm J}}}
\providecommand{\rj}{\ensuremath{\,R_{\rm J}}}
\providecommand{\me}{\ensuremath{\,M_{\rm E}}}

\providecommand{\fave}{\langle F \rangle}
\providecommand{\fluxcgs}{10$^9$ erg s$^{-1}$ cm$^{-2}$}

\submitjournal{AJ}

\begin{document}
\nolinenumbers

\title{Confirmation and Characterization of the Eccentric, Warm Jupiter TIC~393818343~b with a Network of Citizen Scientists}

\correspondingauthor{Lauren A. Sgro}
\email{lsgro@seti.org}



\author[0000-0001-6629-5399]{Lauren A. Sgro}
\affiliation{SETI Institute, Carl Sagan Center, 339 Bernardo Ave, Suite 200, Mountain View, CA 94043, USA}

\author[0000-0002-4297-5506]{Paul A.\ Dalba}
\altaffiliation{Heising-Simons 51 Pegasi b Postdoctoral Fellow}
\affiliation{SETI Institute, Carl Sagan Center, 339 Bernardo Ave, Suite 200, Mountain View, CA 94043, USA}
\affiliation{Department of Astronomy and Astrophysics, University of California, Santa Cruz, CA 95064, USA}

\author[0000-0002-0792-3719]{Thomas M. Esposito}
\affiliation{SETI Institute, Carl Sagan Center, 339 Bernardo Ave, Suite 200, Mountain View, CA 94043, USA}
\affiliation{Unistellar, 5 all\'ee Marcel Leclerc, b\^{a}timent B, Marseille, 13008, France}
\affiliation{Department of Astronomy, University of California, Berkeley, CA 94720, USA}

\author[0000-0001-7016-7277]{Franck Marchis}
\affiliation{SETI Institute, Carl Sagan Center, 339 Bernardo Ave, Suite 200, Mountain View, CA 94043, USA}
\affiliation{Unistellar, 5 all\'ee Marcel Leclerc, b\^{a}timent B, Marseille, 13008, France}


\author[0000-0003-2313-467X]{Diana Dragomir} 
\affiliation{Department of Physics \& Astronomy, University of New Mexico, 1919 Lomas Blvd NE, Albuquerque, NM 87131, USA}

\author[0000-0001-6213-8804]{Steven Villanueva Jr.} 
\altaffiliation{NASA Postdoctoral Program Fellow}
\affiliation{NASA Goddard Space Flight Center, 8800 Greenbelt Road, Greenbelt, MD 20771, USA}

\author[0000-0003-3504-5316]{Benjamin Fulton} 
\affiliation{NASA Exoplanet Science Institute/Caltech-IPAC, MC 314-6, 1200 E. California Blvd., Pasadena, CA 91125, USA}


\author[0000-0002-3278-9590]{Mario Billiani}
\affiliation{Unistellar Network, Vienna, Austria}
\altaffiliation{Citizen Scientist}
\author{Margaret Loose}
\affiliation{Unistellar Network, Santa Fe, New Mexico, USA}
\altaffiliation{Citizen Scientist}
\author[0000-0002-5105-635X]{Nicola Meneghelli}
\affiliation{Unistellar Network, Zevio, Verona, Italy}
\altaffiliation{Citizen Scientist}
\author[0000-0002-1240-6580]{Darren Rivett}
\affiliation{Unistellar Network, Lake Macquarie, NSW, Australia}
\altaffiliation{Citizen Scientist}
\author[0009-0003-6483-0433]{Fadi Saibi}
\affiliation{Unistellar Network, Sunnyvale, CA, USA}
\altaffiliation{Citizen Scientist}
\author[0009-0006-8494-5408]{Sophie Saibi}
\affiliation{Unistellar Network, Sunnyvale, CA, USA}
\altaffiliation{Citizen Scientist}


\author{Bryan Martin}
\affiliation{Exoplanet Watch, Great Falls, MT, USA}
\altaffiliation{Citizen Scientist}
\author[0000-0003-3559-0840]{Georgios Lekkas}
\affiliation{Exoplanet Watch}
\altaffiliation{Citizen Scientist}
\author{Daniel Zaharevitz}
\affiliation{Exoplanet Watch, Frederick, MD, USA}
\altaffiliation{Citizen Scientist}
\author[0000-0001-7547-0398]{Robert T. Zellem}
\affiliation{Jet Propulsion Laboratory, California Institute of Technology, 4800 Oak Grove Drive, Pasadena, CA 91109, USA}
\affiliation{Exoplanet Watch}

\author[0000-0002-0654-4442]{Ivan A.\ Terentev} 
\affiliation{Visual Survey Group, Petrozavodsk, Russia}
\altaffiliation{Citizen Scientist}

\author[0000-0002-5665-1879]{Robert Gagliano} 
\affiliation{Visual Survey Group, Glendale, Arizona, USA}
\altaffiliation{Citizen Scientist}

\author[0000-0003-3988-3245]{Thomas Lee Jacobs} 
\affiliation{Visual Survey Group, Bellevue, WA, USA}
\altaffiliation{Citizen Scientist}

\author[0000-0002-2607-138X]{Martti H.\ Kristiansen} 
\affiliation{Brorfelde Observatory, Observator Gyldenkernes Vej 7, DK-4340 T\o{}ll\o{}se, Denmark}

\author[0000-0002-8527-2114]{Daryll M.\ LaCourse} 
\affiliation{Visual Survey Group, Marysville, WA, USA}
\altaffiliation{Citizen Scientist}

\author{Mark Omohundro} 
\affiliation{Department of Physics, University of Oxford, Denys Wilkinson Building, Keble Road, Oxford, OX13RH, UK}
\altaffiliation{Citizen Scientist}

\author[0000-0002-1637-2189]{Hans M.\ Schwengeler} 
\affiliation{Visual Survey Group, Bottmingen, Switzerland}
\altaffiliation{Citizen Scientist}


\begin{abstract}
\nolinenumbers
NASA's \textit{Transiting Exoplanet Survey Satellite} (TESS) has identified over 7,000 candidate exoplanets via the transit method, with gas giants among the most readily detected due to their large radii. Even so, long intervals between TESS observations for much of the sky lead to candidates for which only a single transit is detected in one TESS sector, leaving those candidate exoplanets with unconstrained orbital periods. Here, we confirm the planetary nature of \planet, originally identified via a single TESS transit, using radial velocity data and ground-based photometric observations from citizen scientists with the Unistellar Network and Exoplanet Watch. We determine a period of $P$ = 16.24921 $\substack{+0.00010 \\ -0.00011}$ days, a mass $M_{P}$ = 4.34 $\pm$ 0.15 $M_{J}$, and semi-major axis $a$ = 0.1291 $\substack{+0.0021 \\ -0.0022}$ au, placing \planet~in the ``warm Jupiter'' population of exoplanets. With an eccentricity $e$ = 0.6058 $\pm$ 0.0023, \planet~is the most eccentric warm Jupiter to be discovered by TESS orbiting less than 0.15 au from its host star and therefore an excellent candidate for follow-up, as it may inform our future understanding of how hot and warm Jupiter populations are linked. 

\end{abstract}

\section{Introduction}\label{sec:intro}
In September 2023, the NASA Exoplanet Archive (NEA) announced the 5,500th confirmed exoplanet \footnote{https://exoplanetarchive.ipac.caltech.edu/}. This growing number is far from the first exoplanet discovery of 51~Peg~b \citep{Mayor1995}, which we now know as a hot Jupiter. Hot Jupiters are typically categorized as having $\geq0.25$ Jupiter masses and orbital periods of $\leq$ 10 days, and are among the most easily detected via the transit method due to their large radii, small semi-major axes, and frequent transits. However, a growing population of discovered warm Jupiters ($P$ $\approx$ 10-200 days, $a$ = 0.1-0.5 au) has informed our understanding of hot Jupiter formation, as it is expected that hot Jupiters reach their orbits via migration from outside the snowline \citep{Dawson2018}. Migration pathways are favored over in situ formation in part due to the large amount of solids required, ${\sim}10$~$\me$, for core accretion \citep{rafikov2006atmospheres} and the lack of observational evidence for formation by gravitational instability, although there is active work investigating this latter scenario \citep{hall2020predicting,paneque2021spiral}. Therefore, this intermediate population of warm Jupiters may be progenitors of hot Jupiters and links to our solar system's distant, cold Jupiter.

Two primary migration pathways have been proposed that may produce hot Jupiters: disk migration, in which an exchange of angular momentum with the surrounding protoplanetary disk causes the planet's migration \citep{Goldreich1980,Baruteau2014}, and high eccentricity (tidal) migration, where a gas giant on an eccentric orbit undergoes tidal dissipation due to close periastron passage with the host star, reducing the planets's semi-major axis and therefore circularizing the orbit (\citealt{Rasio1996,Wu2003}, see \citealt{Dawson2018} for a recent review). To experience tidal migration, a warm Jupiter must have a particular combination of eccentricity and semi-major axis that brings it sufficiently close to its host. Multiple mechanisms for the origin of this high eccentricity have been proposed e.g., Kozai-Lidov oscillations \citep{Kozai1962,Lidov1962,Naoz2016}, secular interactions, \citep{Wu2011,petrovich2016}, and planet-planet scattering \citep{Rasio1996,Ford2006a,Chatterjee2008,Nagasawa2008}. However, only three warm Jupiters are known that exhibit an eccentricity large enough to eventually join the hot Jupiter population via tidal migration, implying an entire population of hot Jupiters with only suspected origins (\citealt{Naef2001,Dong2021}, Gupta et al. 2024, in prep.). The population of eccentric warm Jupiters with unknown fates is important to study and add to, as their evolution history may contain an important piece to the planetary formation puzzle.

With the conclusion of the Kepler and K2 missions, NASA's \textit{Transiting Exoplanet Survey Satellite} \citep[TESS,][]{Ricker2015} has become the most promising observatory with which to detect warm gas giants. There are already 27 confirmed planets in the NASA Exoplanet Archive \citep{ps}\footnote{Accessed on 2024-05-22 at 20:00 UTC.} representative of this population ($a$ = 0.1 -- 0.5 au, $M$ $\geq0.25$ $M_{J}$, and $P$ ranging from 10 to 98 days), all discovered by TESS despite its recent launch in 2018. However, TESS only observes each sector of the sky for an average of 27.4 days, leaving a population of TESS candidates for which only a single transit has been detected. Such candidates have poorly constrained periods and must rely on follow-up photometric and radial velocity (RV) efforts to confirm system properties. One such way to detect subsequent transits is via photometry gathered from the citizen science community. Enlisting such global networks mitigates the limitations faced by professional telescope facilities, such as visibility and weather restrictions of a single location and finite available time. The exoplanet community is already successfully working with these networks through programs such as the Unistellar Network \citep{peluso2023unistellar}, Exoplanet Watch \citep{Zellem2020}, and ExoClock \citep{Kokori_2022,Kokori_2023}. Here we add to the quickly growing list of citizen science successes in exoplanet follow-up.

We present the confirmation of a high-eccentricity ($e$ = 0.6058 $\pm$ 0.0023) warm Jupiter TIC 393818343 b with $P$ = 16.24921 $\substack{+0.00010 \\ -0.00011}$ days using both ground-based transit and RV data. We first identified a single transit signature via visual inspection in Sector 55 TESS data as part of the Visual Survey Group \citep{kristiansen2022visual}. Follow-up observations and analyses with Lick Observatory, the Unistellar Network of Citizen Astronomers, and Exoplanet Watch confirm the existence of this exoplanet. We present our findings in the following order: in Section~\ref{sec:obs} we present our observations of \planet, in Section~\ref{sec:methods} we detail our methods of modeling the system in order to plan photometric observations and confirm the planet, in Section~\ref{sec:res} we discuss our results, and in Section~\ref{sec:disc} we discuss the implications of this discovery and potential future observations. Finally, in Section~\ref{sec:sum} we give an overview of our findings.

\section{Observations\label{sec:obs}}

\subsection{TESS Photometry}\label{sec:tess}

\host was observed by TESS at a 2-minute cadence during its extended mission in Sector 55 from 2022 August 5 to September 1. We use the image data as processed by the Science Processing Operations Center \citep[SPOC,][]{Jenkins2016} at NASA Ames Research Center \citep{https://doi.org/10.17909/t9-nmc8-f686}. Although no evidence of two or more transits was found by the Quick Look Pipeline, additional analyses were carried out by the Visual Survey Group (VSG). Using the \texttt{LcTools} software package \citep{Schmitt2019,schmitt2021lctools}, this group of professional and citizen scientists detected a single transit in TESS data taken on 2022 August 19, prompting a Doppler monitoring follow-up campaign. We refer to the candidate exoplanet that caused this transit signature as \toi. The TESS observation, along with all other photometric observations used in this work, are detailed in Table~\ref{tab:obs}. The candidate \toi~was further categorized as a TESS object of interest, TOI 6883.01, on 2024 February 1 \citep{Conzo_2024}. 

\startlongtable
\begin{deluxetable}{ccccc}
\tabletypesize{\footnotesize}
\tablecaption{Summary of Photometric Observations \label{tab:obs}}
\tablehead{
  \colhead{UTC Date} & 
  \colhead{Transit Epoch} & 
  \colhead{Observer} & 
  \colhead{Telescope} &
  \colhead{Filter}
  }
\startdata
2022 Aug 19 &  0 & \nodata          & TESS   & TESS \\
2023 May 23 & 17 & N. Meneghelli & eVscope-A & Clear\\
2023 May 23 & 17 & M. Billiani   & eVscope-B & Clear\\
2023 Jun 24 & 19 & B. Martin     & 12-in Dall-Kirkham & I\\
2023 Jun 24 & 19 & D. Rivett     & eVscope-C & Clear\\
2023 Jun 24 & 19 & F. \& S. Saibi & eVscope-D & Clear\\
2023 Jun 24 & 19 & M. Loose      & eVscope-E & Clear\\
\enddata
\end{deluxetable}

\subsection{HIRES \& Levy Spectroscopy}\label{sec:apf}

After the discovery of the Sector 55 TESS transit by the VSG, we obtained a high signal-to-noise (S/N > 200) template spectrum of \host taken with the High Resolution Echelle Spectrometer \citep[HIRES,][]{Vogt1994} at W. M. Keck Observatory without the HIRES iodine cell.  We analyzed this spectrum using the synthetic spectral matching tool \texttt{SpecMatch} \citep{Petigura2017b} to gather basic stellar parameters as priors in our global system modeling including effective temperature $T_{eff}$, surface gravity $logg$, metallicity [Fe/H], and projected rotational velocity \textit{vsini}. After confirming via the HIRES data that \host was an appropriate target for Doppler spectroscopy, we proceeded with a follow-up campaign. 

\begin{figure}
    \centering
    \includegraphics[width=\columnwidth]{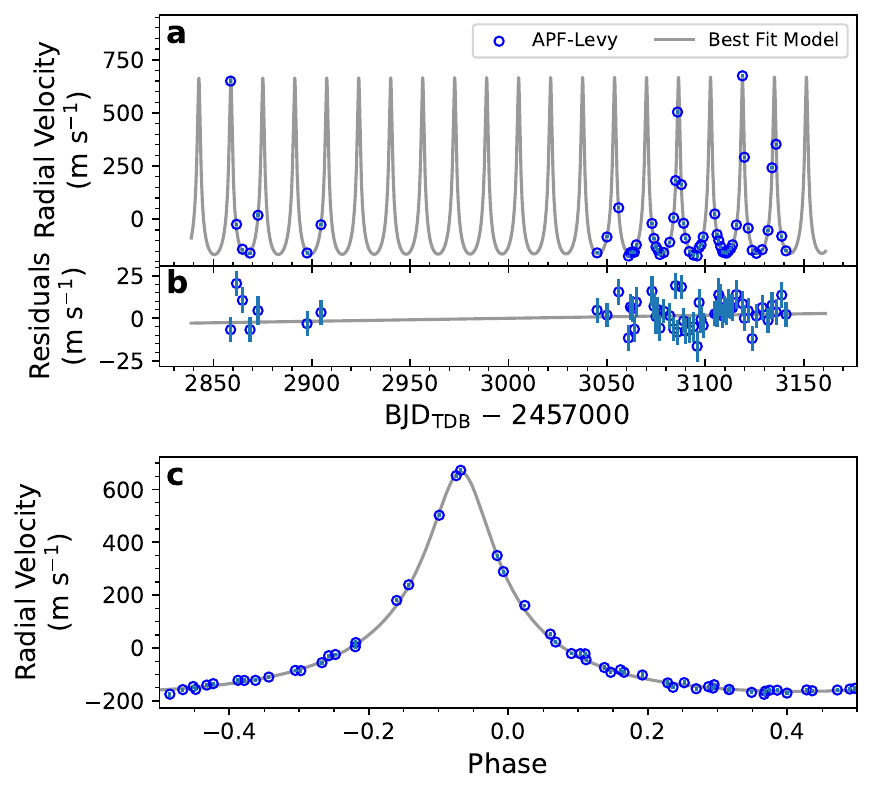}
    \caption{APF-Levy RVs of TIC 393818343. Panel a is the time series. Panel c is phase-folded on the best-fit ephemeris. Panel b shows that an additional long-term RV trend may be present, although from this data set alone it is statistically consistent with no trend.}
    \label{fig:rv}
\end{figure}

From 2022 October to 2023 July, we initiated a Doppler monitoring campaign of \host using the Levy Spectrograph on the Automated Planet Finder (APF) telescope at Lick Observatory in Mt Hamilton, California \citep{Radovan2014, Vogt2014}. The Levy Spectrograph is a high resolution slit-fed optical echelle spectrometer with R $\approx$ 114,000 \citep{Radovan2010} and an iodine cell for precise wavelength calibration \citep{Butler1996}. Over this observing period we gathered 56 RV measurements with uncertainties ranging from 3.6 -- 7.8 $ms^{-1}$. For each spectrum, we deduced RV using a forward modeling procedure \citep{Butler1996, Fulton2015a} in which we used the Keck HIRES spectrum as a template. Obtaining a template spectrum of the necessary quality was more efficient with Keck-HIRES due to the aperture, which is 10 m as opposed to APF's 2.4-m aperture. All RV measurements from APF-Levy are presented in Table~\ref{tab:rv_apf} and time series RVs are shown in Figure~\ref{fig:rv}. 

We combined these RV measurements with the TESS photometry using \texttt{EXOFASTv2} \citep{Eastman2013,Eastman2019} to predict future transit windows, which enabled a ground-based photometric campaign for \toi~ transits. The follow-up photometry is described in Section~\ref{sec:eV} and ~\ref{sec:EW}, while system modeling is described in Section~\ref{sec:methods}. 
\startlongtable
\begin{deluxetable}{cc}
\tabletypesize{\footnotesize}
\tablecaption{RV Measurements of \host\ From APF-Levy. \label{tab:rv_apf}}
\tablehead{
  \colhead{BJD$_{\rm TDB}$} & 
  \colhead{RV (m s$^{-1}$)}
  }
\startdata
$2459858.773894$ & $679.6\pm4.0$ \\
$2459861.776856$ & $6.4\pm4.3$ \\
$2459864.782496$ & $-110.4\pm4.4$ \\
$2459868.654536$ & $-128.9\pm3.9$ \\
$2459872.684238$ & $48.9\pm5.8$ \\
$2459897.626448$ & $-128.3\pm3.9$ \\
$2459904.706206$ & $4.5\pm3.8$ \\
$2460044.993459$ & $-127.8\pm3.7$ \\
$2460050.016955$ & $-53.0\pm3.6$ \\
$2460055.956763$ & $84.3\pm4.3$ \\
$2460060.938359$ & $-143.0\pm4.1$ \\
$2460061.929630$ & $-126.5\pm4.6$ \\
$2460062.935416$ & $-123.3\pm4.5$ \\
$2460063.943645$ & $-124.6\pm4.0$ \\
$2460064.929811$ & $-90.1\pm6.0$ \\
$2460072.904212$ & $10.6\pm6.5$ \\
$2460073.922257$ & $-59.6\pm4.5$ \\
$2460074.940567$ & $-99.8\pm7.8$ \\
$2460075.891244$ & $-114.0\pm4.1$ \\
$2460076.895873$ & $-135.5\pm4.5$ \\
$2460078.891853$ & $-126.6\pm3.7$ \\
$2460081.896280$ & $-77.8\pm4.4$ \\
$2460083.911727$ & $37.0\pm3.6$ \\
$2460084.875597$ & $211.9\pm4.6$ \\
$2460085.866808$ & $534.0\pm4.0$ \\
$2460087.860502$ & $193.1\pm3.6$ \\
$2460088.945967$ & $11.3\pm3.9$ \\
$2460089.862875$ & $-59.5\pm4.2$ \\
$2460091.854226$ & $-121.5\pm4.9$ \\
$2460093.968028$ & $-138.0\pm6.5$ \\
$2460095.840985$ & $-142.0\pm6.8$ \\
$2460096.851126$ & $-102.3\pm4.5$ \\
$2460097.835640$ & $-89.7\pm4.8$ \\
$2460098.891130$ & $-53.1\pm3.9$ \\
$2460104.828737$ & $55.1\pm4.1$ \\
$2460105.964703$ & $-41.2\pm5.0$ \\
$2460106.848495$ & $-69.4\pm7.7$ \\
$2460107.823347$ & $-98.4\pm7.2$ \\
$2460108.871167$ & $-123.8\pm3.8$ \\
$2460109.810574$ & $-126.8\pm4.7$ \\
$2460110.805436$ & $-128.9\pm4.8$ \\
$2460111.796328$ & $-119.6\pm5.9$ \\
$2460112.950157$ & $-107.5\pm4.0$ \\
$2460113.824174$ & $-89.9\pm5.0$ \\
$2460115.789096$ & $3.5\pm5.6$ \\
$2460118.864396$ & $704.4\pm4.2$ \\
$2460119.859491$ & $321.7\pm4.4$ \\
$2460121.786157$ & $-11.4\pm4.2$ \\
$2460123.811574$ & $-115.6\pm4.1$ \\
$2460125.961876$ & $-130.1\pm3.7$ \\
$2460128.882675$ & $-112.7\pm4.3$ \\
$2460131.882008$ & $-22.3\pm3.7$ \\
$2460133.902699$ & $272.0\pm5.0$ \\
$2460135.963745$ & $382.6\pm4.2$ \\
$2460138.837038$ & $-49.3\pm4.2$ \\
$2460140.992830$ & $-118.3\pm3.8$ \\
\enddata
\end{deluxetable}

\subsection{Unistellar Network Observations}\label{sec:eV}

Using the Unistellar Network of citizen scientists, we obtained photometry of \host in accordance with our predicted transit windows calculated as described in Section~\ref{sec:methods}. This network consists of observers around the world who operate mobile, 114-mm Newtonian imaging telescopes, called eVscopes, produced by Unistellar \citep{Marchis2020}. The two models of eVscope both utilize a SONY CMOS sensor, either IMX224 or IMX347, with a 28$^\prime$ $\times$ 37$^\prime$ or 34$^\prime$ $\times$ 47$^\prime$ field of view, respectively. See \citet{peluso2023unistellar} for an in-depth review of the Unistellar Network Exoplanet program. 

On 2023 May 23, two citizen scientists observed \host during a predicted transit window from Austria and Italy. This prompted another follow-up campaign (see Section~\ref{sec:methods}) on 2023 June 24 that included three participating Unistellar Network telescopes located in California, U.S., and Australia.

For each eVscope observation, which varied in duration, off-target or saturated frames were removed from the data set. All other images were dark subtracted and plate solved. Calibrated frames were then aligned and averaged together up to a total integration time of two minutes each to increase the star S/N and reduce computation time in subsequent steps. Following the procedure of \citet{Dalba2016} and \citet{Dalba2017a}, differential photometry was performed on each dataset using an optimized combination of up to ten reference stars. The pool of potential reference stars is created by identifying all stars in each image (other than the target), measuring their fluxes, and removing those which are saturated or do not appear in $\ge50$\% of the frames. We then select up to ten of the brightest stars and generate individual reference light curves for each of them. Next, we generate ``composite reference star'' light curves from every unique combination of those individual reference light curves, where the combination is a median across stars (not time). As a result, we produce up to 1,023 composite reference star light curves with which to normalize the target star's flux. The best normalized target light curve from that group is selected as the one with the smallest residuals in the out-of-transit measurements when compared to a flat model. An optimal aperture size is deduced for each observation by fitting a 2-D Gaussian function to the target across all images in the dataset and calculating a weighted average standard deviation across all frames, where the weights are inversely proportional to the variance. For the only Unistellar dataset crossing ingress, we tested a variety of aperture sizes from 14$^{\prime\prime}$ - 24$^{\prime\prime}$ to ensure that the ingress signal was not introduced by a varying PSF as seeing conditions changed. With this additional analysis, we confirmed that the ingress signal was present in all iterations of the photometry and that therefore our detection was robust. We do not detrend for airmass for any of the four datasets that contain in-transit data since they do not have both pre- and post-transit data points.  

We note that there is a spatially coincident, unbound star, Gaia DR3 4233021429372256512, seen 6.5$^{\prime\prime}$ from \host that can appear blended with the target in ground-based images. We include this nearby source in the target aperture when performing differential photometry and discuss its potential impact in Section~\ref{sec:res}.

\subsection{Exoplanet Watch Observations}\label{sec:EW}

Several citizen scientists associated with Exoplanet Watch observed during both campaigns with telescopes other than eVscopes. Two observations on 2023 May 23 were taken outside of the transit window and therefore contained only baseline data. We do not include these observations in Table~\ref{tab:obs} or in subsequent modeling. However, a 4.5-hour observation taken 2024 June 24 from Utah, U.S., did cover the predicted transit window and is detailed in Table~\ref{tab:obs}. The observer used a Johnson I filter on a 305-mm Dall-Kirkham telescope equipped with a ZWO ASI 294mm CMOS camera (FOV 63$^{\prime}$ $\times$ 31$^{\prime}$) and an exposure time of 90 s per image. The data were $2\times2$ binned and calibrated with dark, flat, and bias frames using $\texttt{EXOTIC}$ \citep{Zellem2020}. $\texttt{EXOTIC}$ was also used by the observer to perform plate solving and differential photometry with a 19.5$^{\prime\prime}$-diameter target aperture. 


\section{Methods}\label{sec:methods}

In this section we describe the global system modeling, the two follow-up photometric campaigns, and the subsequent modeling once additional data were received.

\subsection{Initial System Modeling}
Using the single TESS transit and the subsequent APF-Levy RV observations, we initially modeled the stellar and planetary parameters for the \host system via the \texttt{EXOFASTv2} suite \citep{Eastman2013, Eastman2019}, which jointly considers photometry and RV data when modeling a system. Preliminary modeling included TESS photometry and APF-Levy RV data through May of 2023 along with archival broadband photometry from Gaia DR2 \citep{Gaia2016, Gaia2018}, the \textit{Two Micron All Sky Survey} \citep[2MASS,][]{Cutri2003}, and the \textit{Wide-field Infrared Survey Explorer} \citep[WISE,][]{Cutri2014}. These archival datasets, as well as Gaia parallax measurements, were used by \texttt{EXOFASTv2} to fit spectral energy distributions (SEDs) from MIST stellar evolutionary models \citep{Dotter2016} to gather host star properties. 

Based on the \texttt{Specmatch} analysis of the HIRES \host template spectrum, we placed normal host-star priors on $T_{eff}$ and [Fe/H], as well as on parallax using the Gaia DR3 parallax data corrected for the zero-point offset \citep{2023gaia,Lindegren2021}. We also placed a uniform prior on extinction in the V-band ($A_{V}\leq0.2455$) derived from galactic dust maps \citep{Schlafly2011}. These informative priors are listed at the top of Table~\ref{tab:exofastparams}, and noninformative priors used by \texttt{EXOFASTv2} can be found in Table 3 of \citet{Eastman2019}. 

We consider the fit\textemdash including archival, TESS, and RV data\textemdash converged in accordance with the default \texttt{EXOFASTv2} statistics, where the number of independent draws of the underlying posterior probability distribution is $T_{z}$ > 1000 \citep{Ford2006a}, and the Gelman-Rubin statistic \citep{Gelman1992} < 1.01 for every fitted parameter. This \texttt{EXOFASTv2} analysis initially returned a posterior period with 68\% confidence of 16.2511 $\pm$ 0.0046 days and duration of 4.2 hours, predicting the next mid-transit time to be $T_{o}$ = 2460087.50889 $\pm$ 0.08 BJD\_TDB, or UTC 2023-05-23 00:09:19. 

\subsection{Initial Photometric Campaign \& Subsequent Modeling}
We refer to this particular transit window with mid-transit time $T_{o}$ = 2460087.50889 $\pm$ 0.08 BJD\_TDB as Epoch 17, where Epoch 0 is defined as the TESS Sector 55 transit. We prompted the Unistellar Network and Exoplanet Watch to observe this 2$\sigma$ window plus half the transit time, a 12-hour span during which we obtained four multi-hour observations. Two of these observations consisted only of baseline data, thus they are not included in our tables or figures (see Section 2.4).

Two of the Epoch 17 observations from the Unistellar Network spanned the projected egress time. For each light curve we use a custom transit fitting routine adapted for eVscope data that uses the \texttt{pycheops} Python package and a least-squares minimization to model the transit \citep{peluso2023unistellar}. Due to the partial nature of the individual light curves, the transit duration and period were fixed to those values determined by \texttt{EXOFASTv2}, while the mid-transit time and depth were allowed to vary within 2$\sigma$ and 40\% of the expected value, respectively. Our custom routine then uses the Schwarz criterion rankings from \citet{Kass1995} to determine if the fitted transit is statistically favored over a flat line, among other tests that ensure the model is reasonable. As a result of the analyses included in this routine, observation during Epoch 17 resulted in a preliminary detection of egress. 

The two Epoch 17 datasets containing egress signatures were then combined into the \texttt{EXOFASTv2} fit with the previous TESS photometry and continued RV monitoring data through June of 2023 in order to refine the period to $P$ = 16.24927 $\pm$ 0.0046 days.

\subsection{Confirmation Photometric Campaign}
 We initiated a confirmation campaign to observe a predicted transit with mid-transit time at $T_{o}$ = 2460119.97632 $\pm$ 0.08 BJD\_TDB, or 2023-06-24 11:18:53 UTC, two periods after our initial endeavor. We refer to this transit window, which spanned eight hours to allow for two-hour baseline observations in addition to the expected transit duration of $\sim$4~hours, as Epoch 19. 

\begin{figure}
    \centering
    \includegraphics[width=\columnwidth]{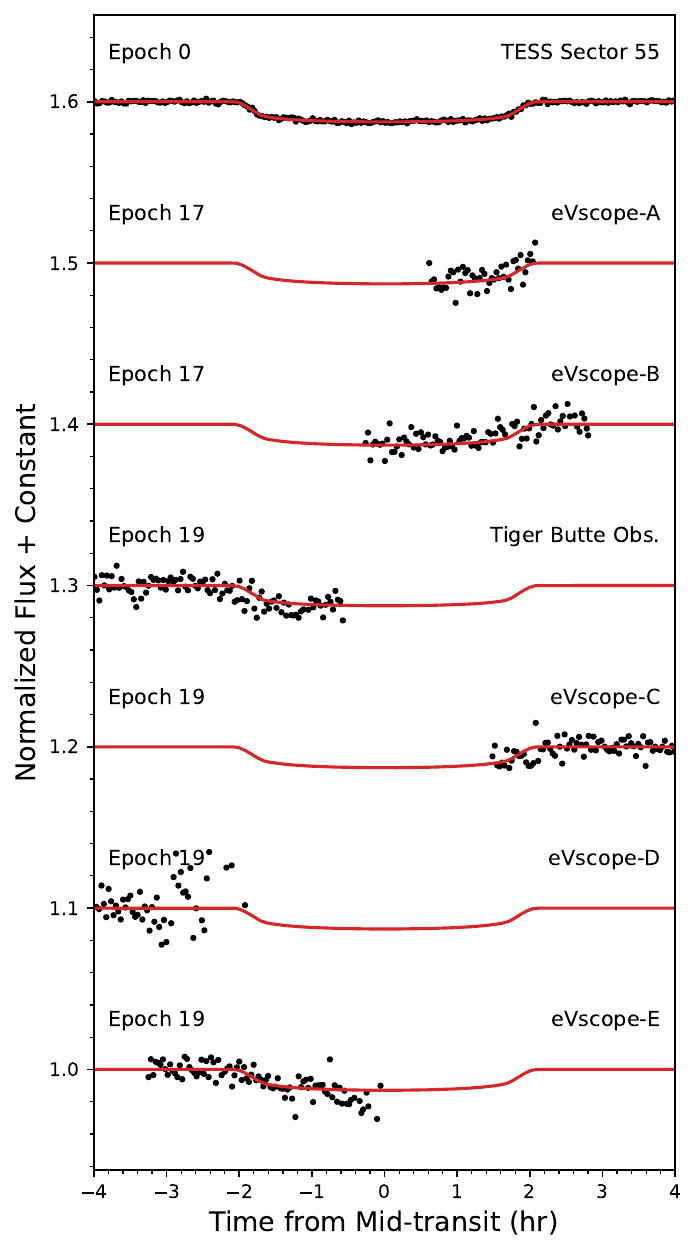}
    \caption{Individual transit observations of \planet\ and the best-fit model from \texttt{EXOFASTv2} shown in red. The names of observers and relevant epochs are listed in Table~\ref{tab:obs}.}
    \label{fig:transits}
\end{figure}

\begin{figure*}
    \centering
    \includegraphics[width=\textwidth]{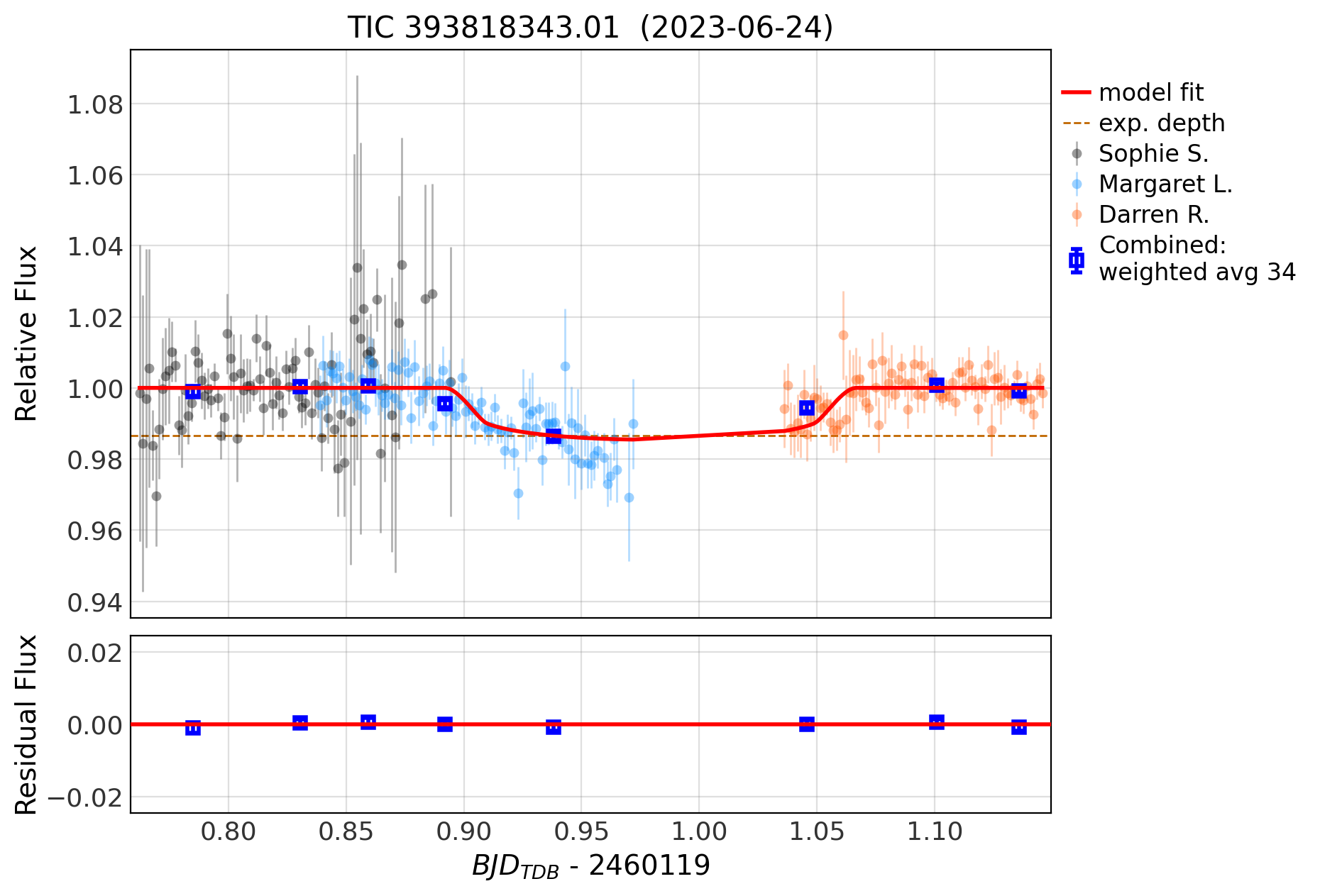}
    \caption{Light curve of observations taken by Unistellar Network observers during the second photometric confirmation campaign (Epoch 19, described in Section 3.3). Note that observer Margaret L.'s data were affected by fog near the expected mid-transit time.}
    \label{fig:uni}
\end{figure*}

Four observers took part in our Epoch 19 confirmation campaign, with their data covering predicted ingress and egress. Light curves from all photometric campaign epochs are presented individually in Figure~\ref{fig:transits}. The Epoch 19 campaign was successful, as the transit was detected in a joint light curve combining the three Unistellar eVscope data sets, shown in Figure~\ref{fig:uni}. Another observation from Exoplanet Watch also detected ingress independently. These four datasets were then combined with the complete APF RV dataset once Doppler monitoring concluded in 2023 July, as well as with the ground-based Epoch 17, TESS Epoch 0, and archival photometry, to be used as input for a final \texttt{EXOFASTv2} analysis following the process described in Section 3.1. We discuss the results of this analysis in Section~\ref{sec:res}.

\startlongtable
\begin{deluxetable*}{lcccc}
\tablecaption{Median values and 68\% confidence interval for TIC393818343}
\tablehead{\colhead{~~~Parameter} & \colhead{Units} & \multicolumn{3}{c}{Values}}
\startdata
\smallskip\\\multicolumn{2}{l}{Informative Priors:}&\smallskip\\
~~~~$T_{\rm eff}$\dotfill &Effective Temperature (K)\dotfill &$\mathcal{N}(5800,100)$\\
~~~~$[{\rm Fe/H}]$\dotfill &Metallicity (dex)\dotfill &$\mathcal{N}(0.32,0.06)$\\
~~~~$\varpi$\dotfill &Parallax (mas)\dotfill &$\mathcal{N}(10.644,0.397)$\\
~~~~$A_V$\dotfill &V-band Extinction (mag))\dotfill &$\mathcal{U}(0,0.2455)$\\
\smallskip\\\multicolumn{2}{l}{Stellar Parameters:}&\smallskip\\
~~~~$M_*$\dotfill &Mass (\msun)\dotfill &$1.082^{+0.055}_{-0.056}$\\
~~~~$R_*$\dotfill &Radius (\rsun)\dotfill &$1.086\pm0.020$\\
~~~~$L_*$\dotfill &Luminosity (\lsun)\dotfill &$1.168^{+0.065}_{-0.063}$\\
~~~~$F_{Bol}$\dotfill &Bolometric Flux (cgs)\dotfill &$0.00000000423\pm0.00000000023$\\
~~~~$\rho_*$\dotfill &Density (cgs)\dotfill &$1.198^{+0.040}_{-0.058}$\\
~~~~$\log{g}$\dotfill &Surface gravity (cgs)\dotfill &$4.402^{+0.013}_{-0.018}$\\
~~~~$T_{\rm eff}$\dotfill &Effective Temperature (K)\dotfill &$5756^{+67}_{-66}$\\
~~~~$[{\rm Fe/H}]$\dotfill &Metallicity (dex)\dotfill &$0.318\pm0.058$\\
~~~~$[{\rm Fe/H}]_{0}$\dotfill &Initial Metallicity$^{1}$ \dotfill &$0.299\pm0.057$\\
~~~~$Age$\dotfill &Age (Gyr)\dotfill &$3.8^{+2.6}_{-2.1}$\\
~~~~$EEP$\dotfill &Equal Evolutionary Phase$^{2}$ \dotfill &$355^{+33}_{-22}$\\
~~~~$A_V$\dotfill &V-band extinction (mag)\dotfill &$0.127^{+0.073}_{-0.076}$\\
~~~~$\sigma_{SED}$\dotfill &SED photometry error scaling \dotfill &$1.28^{+0.49}_{-0.30}$\\
~~~~$\varpi$\dotfill &Parallax (mas)\dotfill &$10.643\pm0.039$\\
~~~~$d$\dotfill &Distance (pc)\dotfill &$93.96^{+0.35}_{-0.34}$\\
~~~~$\dot{\gamma}$\dotfill &RV slope$^{3}$ (m/s/day)\dotfill &$0.001\pm0.015$\\
\smallskip\\\multicolumn{2}{l}{Planetary Parameters:}&b\smallskip\\
~~~~$P$\dotfill &Period (days)\dotfill &$16.24921^{+0.00010}_{-0.00011}$\\
~~~~$R_P$\dotfill &Radius (\rj)\dotfill &$1.087^{+0.023}_{-0.021}$\\
~~~~$M_P$\dotfill &Mass (\mj)\dotfill &$4.34\pm0.15$\\
~~~~$T_C$\dotfill &Time of conjunction (\bjdtdb)\dotfill &$2459811.24020^{+0.00025}_{-0.00024}$\\
~~~~$a$\dotfill &Semi-major axis (AU)\dotfill &$0.1291^{+0.0021}_{-0.0022}$\\
~~~~$i$\dotfill &Inclination (Degrees)\dotfill &$89.57^{+0.30}_{-0.38}$\\
~~~~$e$\dotfill &Eccentricity \dotfill &$0.6058\pm0.0023$\\
~~~~$\omega_*$\dotfill &Argument of Periastron (Degrees)\dotfill &$1.69\pm0.49$\\
~~~~$T_{eq}$\dotfill &Equilibrium temperature$^{4}$ (K)\dotfill &$805.5^{+9.6}_{-9.4}$\\
~~~~$\tau_{\rm circ}$\dotfill &Tidal circularization timescale (Gyr)\dotfill &$26.2^{+2.1}_{-2.6}$\\
~~~~$K$\dotfill &RV semi-amplitude (m/s)\dotfill &$414.8\pm2.8$\\
~~~~$R_P/R_*$\dotfill &Radius of planet in stellar radii \dotfill &$0.10295^{+0.00065}_{-0.00062}$\\
~~~~$a/R_*$\dotfill &Semi-major axis in stellar radii \dotfill &$25.60^{+0.28}_{-0.42}$\\
~~~~$\delta$\dotfill &$\left(R_P/R_*\right)^2$ \dotfill &$0.01060\pm0.00013$\\
~~~~$\delta_{\rm I}$\dotfill &Transit depth in I (fraction)\dotfill &$0.01237^{+0.00038}_{-0.00037}$\\
~~~~$\delta_{\rm Clear}$\dotfill &Transit depth in Clear (fraction)\dotfill &$0.01205\pm0.00025$\\
~~~~$\delta_{\rm TESS}$\dotfill &Transit depth in TESS (fraction)\dotfill &$0.01270\pm0.00022$\\
~~~~$\tau$\dotfill &Ingress/egress transit duration (days)\dotfill &$0.01634^{+0.00067}_{-0.00024}$\\
~~~~$T_{14}$\dotfill &Total transit duration (days)\dotfill &$0.17294^{+0.00089}_{-0.00082}$\\
~~~~$b$\dotfill &Transit impact parameter \dotfill &$0.119^{+0.10}_{-0.083}$\\
~~~~$b_S$\dotfill &Eclipse impact parameter \dotfill &$0.123^{+0.11}_{-0.086}$\\
~~~~$\tau_S$\dotfill &Ingress/egress eclipse duration (days)\dotfill &$0.01698^{+0.00069}_{-0.00033}$\\
~~~~$T_{S,14}$\dotfill &Total eclipse duration (days)\dotfill &$0.1791\pm0.0020$\\
~~~~$\rho_P$\dotfill &Density (cgs)\dotfill &$4.22^{+0.16}_{-0.24}$\\
~~~~$logg_P$\dotfill &Surface gravity \dotfill &$3.962^{+0.010}_{-0.018}$\\
~~~~$\Theta$\dotfill &Safronov Number \dotfill &$0.952^{+0.019}_{-0.020}$\\
~~~~$\fave$\dotfill &Incident Flux (\fluxcgs)\dotfill &$0.0682^{+0.0033}_{-0.0031}$\\
~~~~$T_P$\dotfill &Time of Periastron (\bjdtdb)\dotfill &$2459810.145^{+0.015}_{-0.016}$\\
~~~~$T_S$\dotfill &Time of eclipse (\bjdtdb)\dotfill &$2459808.973\pm0.019$\\
~~~~$e\cos{\omega_*}$\dotfill & \dotfill &$0.6055\pm0.0023$\\
~~~~$e\sin{\omega_*}$\dotfill & \dotfill &$0.0179\pm0.0052$\\
\smallskip\\\multicolumn{2}{l}{Wavelength Parameters:}&Dall-Kirkham (I)& eVscope (Clear) &TESS\smallskip\\
~~~~$u_{1}$\dotfill &linear limb-darkening coeff \dotfill &$0.294\pm0.050$&$0.425\pm0.026$&$0.339\pm0.033$\\
~~~~$u_{2}$\dotfill &quadratic limb-darkening coeff \dotfill &$0.269\pm0.050$&$0.271\pm0.024$&$0.273\pm0.046$\\
\smallskip\\\multicolumn{2}{l}{Telescope Parameters:}&APF-Levy\smallskip\\
~~~~$\gamma_{\rm rel}$\dotfill &Relative RV Offset$^{3}$ (m/s)\dotfill &$33.1\pm1.5$\\
~~~~$\sigma_J$\dotfill &RV Jitter (m/s)\dotfill &$7.23^{+1.1}_{-0.97}$\\
\enddata
\label{tab:exofastparams}
\tablenotetext{}{See Table 3 in \citet{Eastman2019} for a detailed description of all parameters and default, noninformative priors. Informative priors listed as $\mathcal{N}(a,b)$ indicates a normal distribution with mean $a$ and standard deviation $b$, while $\mathcal{U}(a,b)$ indicates a uniform distribution over the interval [$a$, $b$].}
\tablenotetext{1}{The metallicity of the star at birth}
\tablenotetext{2}{Corresponds to static points in a star's evolutionary history. See \S2 in \citet{Dotter2016}.}
\tablenotetext{3}{Reference epoch = 2459999.883362}
\tablenotetext{4}{Assumes no albedo and perfect redistribution}
\end{deluxetable*}

\section{Results}\label{sec:res}

From the best-fit \texttt{EXOFASTv2} solution using all of the available data outlined in Section 3.3, we confirm that TIC 393818343.01, hereafter \planet, is a giant planet with an eccentric orbit at $e$ = 0.6058 $\pm$ 0.0023. The best-fit period $P$ = 16.24921 $\substack{+0.00010 \\ -0.00011}$, in close agreement with the initial estimates, and best-fit mass and radius of $M_{P}$ = 4.34 $\pm$ 0.15 $M_{J}$ and $R_{P}$ = 1.087$\substack{+0.023 \\ -0.021}$ $M_{J}$, respectively, place this planet in the warm Jupiter category of gas giants. Posteriors of the \texttt{EXOFASTv2} modeling are listed in Table~\ref{tab:exofastparams}. 

There is a foreground star, Gaia DR3 4233021429372256512, at a distance of 84.5 pc compared to the target's distance of 93.5 pc \citep{BailerJones2018}, with which the target is blended in ground-based images. We included this star in the target aperture when performing differential photometry, running the risk that the transit depth could be diluted when compared to TESS observations, which are corrected for this dilution, due to the excess baseline flux from the companion. However, we found that the modeled transit depth in Clear, the filter representative of the eVscope, is in agreement with the transit depth measured from the TESS light curve (see Table~\ref{tab:exofastparams}). 

To test this agreement between the corrected TESS and the ground-based data, we assume that the depth of the measured TESS transit is 1.27\% of the host star flux. Note that we used the \texttt{lightkurve} package \citep{Lightkurve2018} as described in \citet{Peluso2024} to reproduce the TESS light curve discovered by the VSG and deduce the transit depth, which uses the TESS PDCSAP flux that has already been corrected for crowding. The crowding metric from the TESS \texttt{TargetPixelFile} for the Sector 55 observation, or \texttt{CROWDSAP}, is 0.77034879, indicating that approximately 23\% of flux in the TESS aperture is from objects other than the host star. If we consider this additional flux in our depth calculation, the resultant "diluted" depth is 1.03\%. This depth is only 0.24\% from the corrected transit depth, which is well within the noise floor of the eVscope observations. We obtain a similar result when considering the "diluted" depth in relation to the difference in magnitude between the host and foreground star (1.38 mag, or 28\% of the host star magnitude in Gaia $G_{rp}$). Therefore, we ignore the effects of dilution from Gaia DR3 4233021429372256512 in our modeling using ground-based data. 

Additionally, the foreground star is identified as a variable in the Gaia DR3 catalog via the \texttt{phot\_variable\_flag}; however, it is unreasonable that the star could cause the observed transit. Gaia DR3 4233021429372256512 is a 10.31 magnitude star in the Gaia \textit{$G_{RP}$} band, while \host \textit{$G_{RP}$} = 8.93 (the TESS bandpass is centered on the traditional Cousins-I band at 786.5 nm, while \textit{$G_{RP}$} is centered at 797 nm). Although the Gaia data indicate a range in \textit{$G_{RP}$} of 0.0163 for the foreground star, the duration of the variability is listed as \texttt{time\_duration\_rp} = 928.9 days. As this value is over 5,000 times the transit duration, it is unlikely that the variability of the foreground star is responsible for the photometric transit signature that was observed and predicted via the \texttt{EXOFASTv2} analysis. 


\section{Discussion}\label{sec:disc}

\subsection{Eccentricity and Tidal Circularization}
\begin{figure}
    \centering
    \includegraphics[width=\columnwidth]{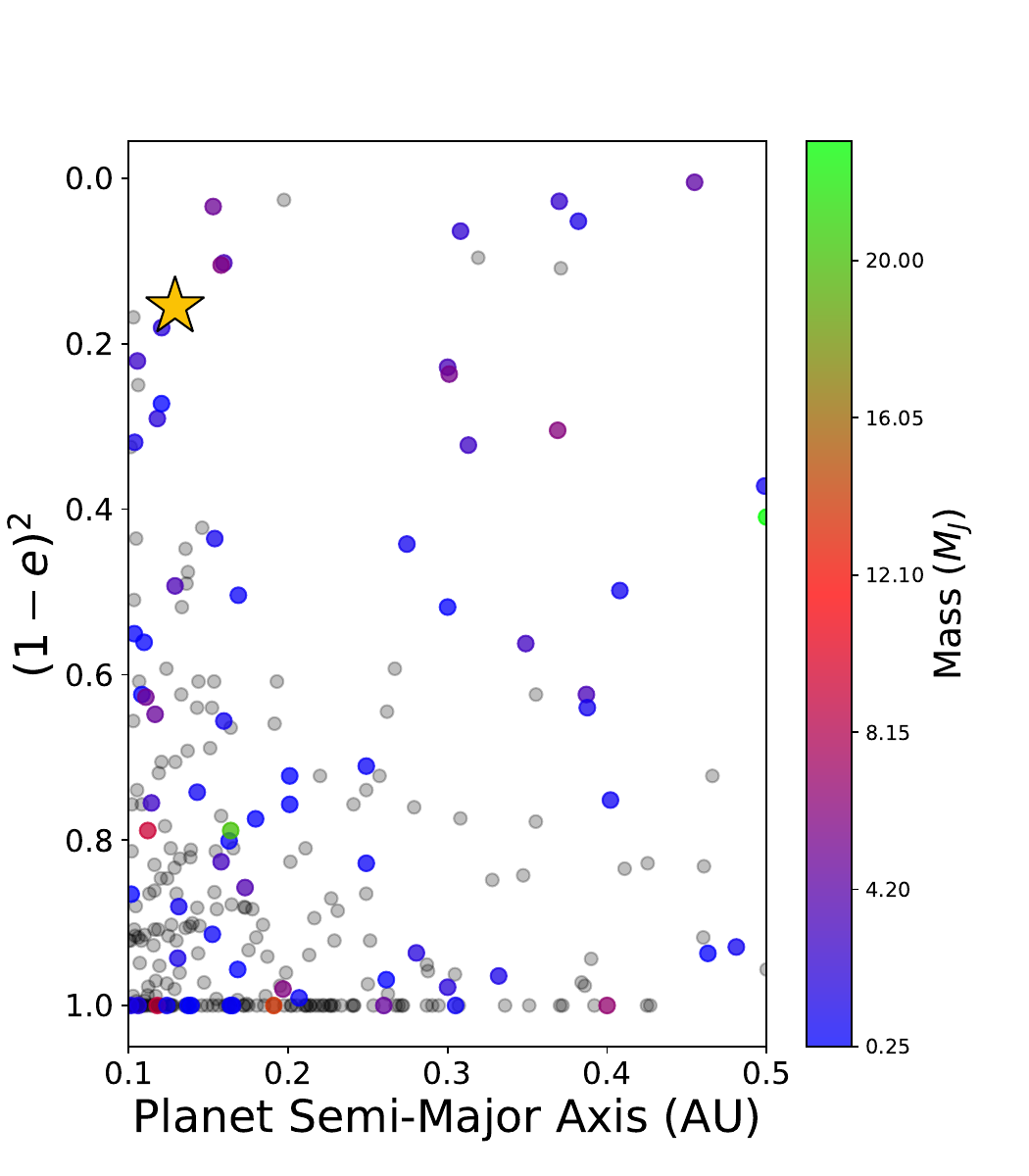}
    \caption{Confirmed planets with 10 < $P$ < 200 days and 0.1 $\leq$ $a$ $\leq$ 0.5 au from the NASA Exoplanet Archive that have available mass measurements. Giant planets with $M_{P}$ $\geq$ 0.25 $M_{J}$ are plotted such that colors denote the mass of the planet, while planets below this mass cutoff are plotted in gray. The yellow star indicates \planet~at $4.34\pm0.15$\mj.}
    \label{fig:params}
\end{figure}

With an eccentricity $e$ = 0.6058 $\pm$ 0.0023, \planet~occupies a sparsely populated orbital parameter space. This feature is evident in Figure~\ref{fig:params}, which compares $e$ and $a$ for a subset of confirmed planets with mass measurements from the NASA Exoplanet Archive \citep{ps}.\footnote{Accessed on 2024-05-22 at 07:30 UTC.} The most eccentric known exoplanets include HD 80606 b \citep[$e$ = 0.93,][]{Naef2001}, TOI-3362 b \citep[$e$ = 0.81,][]{Dong2021} and another warm Jupiter with $e$ = 0.93 (Gupta et al. 2024, in prep.), all of which are expected to tidally circularize before their host star evolves off the main sequence. To compare, we calculate via \texttt{EXOFASTv2} the tidal circularization timescale, $\tau_{\rm circ}$, to be 26.2$\substack{+2.1 \\ -2.6}$ Gyr (see Equation 3 from \citet{Adams2006}, assuming $Q$ = 10\textsuperscript{6}). This timescale is longer than the main sequence lifetime of an early G-type star, which is expected to be $\sim$10 Gyr; however, this lifetime and $\tau_{\rm circ}$ are on the same order of magnitude. The full picture of evolution from warm to hot Jupiter is not fully understood, and \planet~may help complete a picture that suggests tidal migration is more of a spectrum than has previously been understood. Yet despite its sizeable eccentricity, \planet~is not eccentric or close enough to its host to migrate inward and become a hot Jupiter under the stated assumptions in a single-planet system.

Tidal circularization might still be possible if tidal dissipation was expedited, but \planet~is not well-suited for this scenario. \citet{wu2018diffusive} proposes this enhancement is possible through energy exchange between the orbit and the planet's degree-2 fundamental-mode (f-mode) during periastron passage \textit{if} the planet passes within four times the host's tidal radius, $r_{\tau}$ = $R_{P}$($M_{*}$/$M_{P}$)$\textsuperscript{1/3}$. Using the methods from \citet{gupta2023high}, we calculate a periastron distance $a_{peri}$ $\approx$ 7.5$R_{*}$. This value is 2.9 times greater than the maximum distance, 4$r_{\tau}$ = 2.6$R_{*}$, at which \citet{wu2018diffusive} states that f-mode dissipation would be effective to boost tidal dissipation efficiency; therefore, \planet's orbit could not be circularized more quickly through this channel.

Considering that only three planets are expected to tidally circularize into hot Jupiters, there is a disconnect between the observed warm and hot Jupiter populations. However, \citet{petrovich2016} propose a mechanism that can account for much of the hot Jupiter population via secular planet-planet interactions; i.e., in which Kozai-Lidov oscillations occur due to the presence of an outer planetary companion. In this case, warm Jupiters migrate in intervals of high eccentricity but spend more time in low eccentricity phases and are therefore observed largely at lower eccentricities. As an exercise,
 we can consider a case in which \planet~does circularize via this or some other unknown pathway, despite the values of $\tau_{\rm circ}$ and $a_{peri}$. Again following the analysis of \citet{gupta2023high} and assuming that angular momentum is conserved, the final semi-major axis if \planet~were to circularize would be $a_{final}$ = 0.0817 au. Along with the resultant orbital period in this scenario, $P_{final}$ = 8.17 days, this final orbital configuration would land \planet~in the hot Jupiter population. Although it is not currently expected to become a hot Jupiter, \planet~is an interesting case for further research due to its potential to enter this prolific population if an outer companion is discovered.

 Note that the presence of a perturbing outer planet is beyond the scope of this work. Despite no clear evidence from our RV data, the existence of a distant companion cannot be discounted and we leave this investigation to future studies. 

\subsection{Opportunities for Atmospheric Characterization}
Atmospheric characterization via emission and transmission spectroscopy have become more feasible with the launch of the \textit{James Webb Space Telescope} (JWST) and other space observatories. To determine the feasibility of follow-up, we can look at the Emission Spectroscopy Metric (ESM) developed by \citet{Kempton2018} based upon the work of \citet{Zellem_2017}, which is proportional to the S/N that JWST would be expected to achieve with a mid-infrared secondary eclipse detection using MIRI/LRS. Using the properties of \planet~derived from this work, including the dayside temperature as 1.10 $T_{eff}$, the ESM = 57, indicating that this planet is a reasonable candidate for emission spectroscopy. \citet{Kempton2018} recommends a threshold of ESM = 7.5 for candidates, although only terrestrial targets are considered in their analysis for emission spectroscopy. 

However, according to the Transmission Spectroscopy Metric (TSM) \citep{Zellem_2017,Kempton2018}, \planet\xspace is not a suitable target for this type of observation using JWST/NIRISS. For the target, TSM = 20, which is lower than the suggested threshold of 96 for sub Jovians, as \planet\xspace is not large or hot enough to be an ideal transmission spectroscopy target. 

To compare, \citet{mann2023giant} plot various ESM values according to planet mass and temperature for confirmed giant planets ($R > 0.5R_{J}$) to assess follow-up for the cool giant TOI-2010 b. From their analysis, we can see that \planet's ESM falls among the median for all giant planets as well as those of a similar temperature. Emission spectroscopy of \planet~could both expose the temperature structure of its atmosphere via a P-T profile as well as atmospheric abundances, which would add valuable information to the connection between warm and hot Jupiters \citep{Madhusudhan2019, fortney2021hot}. Chemical composition and abundances may indicate whether a gas giant formed in situ or migrated inwards, a key question in the general formation of warm Jupiters and, although emission spectra are available for many hot Jupiters, there is a paucity of this data for their cooler cousins. 

Considering the ESM and an extremely likely occultation of \planet~\citep[$b_{s}$ = 0.123 $\substack{+0.11 \\ -0.086}$, see Equation 7 of][]{Winn2010a}, we believe it is a suitable candidate for emission spectroscopy follow-up. Secondary eclipse timing, $T_{S}$, is available in Table~\ref{tab:exofastparams}. Note that, due to its small semi-major axis $a$, \planet~is not a reasonable target for direct imaging observations. 

\section{Summary}\label{sec:sum}

Here we present a confirmation of the TESS-discovered exoplanet \planet. First identified as a candidate from a single TESS transit in Sector 55, we gathered follow-up RV and photometric observations using both the APF at Lick Observatory and a network of citizen scientists. Our analyses show that \planet~is a warm Jupiter, with $P$ = 16.25 days and $a$ = 0.1291 au, on a highly eccentric orbit, $e$ = 0.6058, around an early G-type star. Its orbital parameters indicate that, if the orbit were to circularize, \planet~would join the hot Jupiter population; however, the timescale for which this would take place via the tidal migration pathway is longer than the main sequence lifetime of the host star. 

We note that this planet is a candidate for atmospheric characterization via emission spectroscopy, although it is not a feasible target for transmission spectroscopy or direct imaging. Future studies of \planet's atmosphere with space telescopes such as JWST or the \textit{Hubble Space Telescope}, could add to the ongoing investigation as the scientific community continues to characterize the warm Jupiter population.

The crucial data from ground-based observations to confirm the planetary nature of \planet~demonstrate the impact of the burgeoning field of citizen science. These data constrained the orbital parameters modeled with \texttt{EXOFASTv2} and confirmed the period derived from the RV analysis. Due to the global nature of the Unistellar Network and Exoplanet Watch, weather effects were mitigated and multiple observations were possible during each campaign, furthering the significance of our detections. This discovery adds to the multiple publications over various fields resulting from such citizen science observations; e.g. \citet{Zellem2020, Pearson2022, Perrocheau2022,graykowski2023light, sgro2023photometry, Peluso2024}. These networks of amateur and professional astronomers will continue to be vital in keeping ephemerides fresh for \planet~and other confirmed exoplanets, as well as in the confirmation of future candidates.

\section{Acknowledgements}

L.A.S. and T.M.E. were partially supported during this work by the NASA Citizen Science Seed Funding Program via grant number 80NSSC22K1130 and the NASA Exoplanets Research Program via grant 80NSSC24K0165. Those NASA grants also support the UNITE (Unistellar Network Investigating TESS Exoplanets) program, under the auspices of which the Unistellar data were collected. P.A.D. acknowledges support by a 51 Pegasi b Postdoctoral Fellowship from the Heising-Simons Foundation. D.D. acknowledges support from the NASA Exoplanet Research Program grant 18-2XRP18\_2-0136. This study also benefited from the kind contributions of the Gordon and Betty Moore Foundation.

The scientific data highlighted in this paper were partially sourced from the Unistellar Network, a collaborative effort between Unistellar and the SETI Institute. We extend our gratitude to Frédéric Gastaldo for financially supporting the foundational phase of the Unistellar Citizen Science project. We express our heartfelt appreciation to the citizen astronomers who shared their invaluable data for these observations. This publication also makes use of data products from Exoplanet Watch, a citizen science project managed by NASA's Jet Propulsion Laboratory on behalf of NASA's Universe of Learning. This work is supported by NASA under award number NNX16AC65A to the Space Telescope Science Institute, in partnership with Caltech/IPAC, Center for Astrophysics|Harvard \& Smithsonian, and NASA Jet Propulsion Laboratory. Part of this research was carried out at the Jet Propulsion Laboratory, California Institute of Technology, under a contract with the National Aeronautics and Space Administration (80NM0018D0004).

We thank Ken and Gloria Levy, who supported the construction of the Levy Spectrometer on the Automated Planet Finder. We thank the University of California and Google for supporting Lick Observatory and the UCO staff for their dedicated work scheduling and operating the telescopes of Lick Observatory. Some of the data presented herein were obtained at the W.~M. Keck Observatory, which was made possible by the generous financial support of the W.~M. Keck Foundation and is operated as a scientific partnership among the California Institute of Technology, the University of California, and NASA. The authors wish to recognize and acknowledge the very significant cultural role and reverence that the summit of Mauna Kea has always had within the indigenous Hawaiian community. We are most fortunate to have the opportunity to conduct observations from this mountain.

This paper includes data collected by the TESS mission. Funding for the TESS mission is provided by the NASA's Science Mission Directorate. This research also made use of the NASA Exoplanet Archive, which is made available by the NASA Exoplanet Science Institute at IPAC, operated by the California Institute of Technology under contract with the National Aeronautics and Space Administration. Additionally, this research made use of the Exoplanet Follow-up Observation Program (ExoFOP; DOI: 10.26134/ExoFOP5) website, which is operated by the California Institute of Technology, under contract with the National Aeronautics and Space Administration under the Exoplanet Exploration Program.



\vspace{5mm}
\facilities{Automated Planet Finder (Levy), Keck:I (HIRES), TESS, Unistellar}\\
\vspace{5mm}

\textit{Data}: All the TESS data used in this paper can be found in MAST:  doi:10.17909/t9-nmc8-f686. All NASA Exoplanet Archive data used in this study can be found on the NASA Exoplanet Archive: 10.26133/NEA12.

\vspace{5mm}
\software{   \textsf{astropy} \citep{astropy2013,astropy2018},
                \texttt{EXOFASTv2} \citep{Eastman2013,Eastman2017,Eastman2019}, 
                \texttt{lightkurve} \citep{Lightkurve2018}, 
                \texttt{SpecMatch} \citep{Petigura2015,Petigura2017b}, 
                \textbf{\texttt{LcTools} \citep{Schmitt2019}}
                }


\bibliography{references}{}
\bibliographystyle{aasjournal}

\end{document}